\documentclass{iau307}
\usepackage{graphicx}
\usepackage{natbib}
\usepackage{url}
\usepackage{dtklogos}
\bibpunct{(}{)}{;}{a}{}{,}

\title[Binarity of the LBV HR\,Car] 
{Binarity of the LBV HR\,Car}

\author[Th.~Rivinius et al.]  
       {Th.~Rivinius$^{1}$ \and H.M.J. Boffin$^{1}$ \and W.J.~de Wit$^{1}$
         \and A.~Mehner$^{1}$ \and Ch.~Martayan$^{1}$ \and S.~Guieu$^{2}$ \and
         J.-B.~Le Bouquin$^{2}$}

\affiliation{$^{1}$ESO - European Organisation for Astronomical Research in the Southern Hemisphere, Chile \\ email: {\tt triviniu@eso.org} \\[\affilskip]
$^{2}$Univ. Grenoble Alpes, CNRS, IPAG, France}

\pubyear{2014}
\volume{307} 
\pagerange{}
\jname{New windows on massive stars: asteroseismology, interferometry, and spectropolarimetry}
\editors{G. Meynet, C. Georgy, J.H. Groh \& Ph. Stee, eds.}

\begin{document}

\maketitle

\begin{abstract}
VLTI/AMBER and VLTI/PIONIER observations of the LBV HR\,Car show an
interferometric signature that could not possibly be explained by an extended
wind, more or less symmetrically distributed around a single object. Instead,
observations both in the Br$\gamma$ line and the $H$-band continuum are best
explained by two point sources (or alternatively one point source and one
slightly extended source) at about 2\,mas separation and a contrast ratio of
about 1:5. These observations establish that HR Car is a binary, but further
interpretation will only be possible with future observations to constrain the
orbit. Under the assumption that the current separation is close to the
maximum one, the orbital period can be estimated to be of the order of 5 years,
similar as in the $\eta$ Car system. This would make HR\,Car the second such
LBV binary.
\end{abstract}

\firstsection 
\section{Introduction}
Luminous Blue Variables (LBVs) are a brief phase in the evolution of massive
stars, but a very important one. The giant eruption remains enigmatic, but the
discovery of the flagship LBV $\eta$\,Car to be a five-year highly eccentric
binary put focus on possible binarity induced mechanism for the giant
outbursts, and prompted binarity searches among LBVs.

So far, however, while several wide LBV binaries were identified, LBV systems
similar to $\eta$\,Car (relatively close and eccentric) have not been found,
with the possible exception of the LBV candidate MWC\,314
(\citealt{2013A&A...559A..16L}; see as well the preliminary summary by
\citealt{2012ASPC..464..293M}). This is rather surprising as it is thought that
given their very high multiplicity rate, more than 70\% of all massive stars
will exchange mass with a companion \citep{2012Sci...337..444S}.

\section{Observations}
\subsection{AMBER/VLTI OHANA Data}

The LBV HR\,Car was observed as part of the OHANA sample (spectrally resolved
3-beam interferometry of Br$\gamma$ with AMBER at the VLTI, see Rivinius et
al., this volume). The OHANA data obtained for HR Car showed a clear and
temporally stable (over the months of observation) phase signature across the
blue part of the emission line, but little to no visibility signature. This
marks a {photocentre displacement of the emission line with respect to the
  continuum}. Such a displacement is hard to explain with the more-or-less
symmetric, but variable wind of a single supergiant star. Ad-hoc explanations
are:
\begin{itemize}
\item {\bf A binary}, and the emission is associated with the secondary, not
  the primary;
\item {\bf A binary}, where part of the emission is formed at the location of
  the secondary, part at the location of the primary, and part in a wind
  collision zone;
\item {\bf A single star} with a dense nebular structure nearby that forms the
  hydrogen emission (possibly ejected in a previous eruption).
\end{itemize}

\subsection{PIONIER Data}

The available OHANA data would not allow further distinction of these
hypotheses, so HR\,Car was observed with PIONIER. {PIONIER is a 4-beam
  interferometric instrument working in the $H$-band continuum}
\citep{2011A&A...535A..67L}, and thus is not as sensitive to nebular
contributions as OHANA observations.

\begin{figure}[t]
\begin{center}
\includegraphics[angle=0,width=\textwidth]{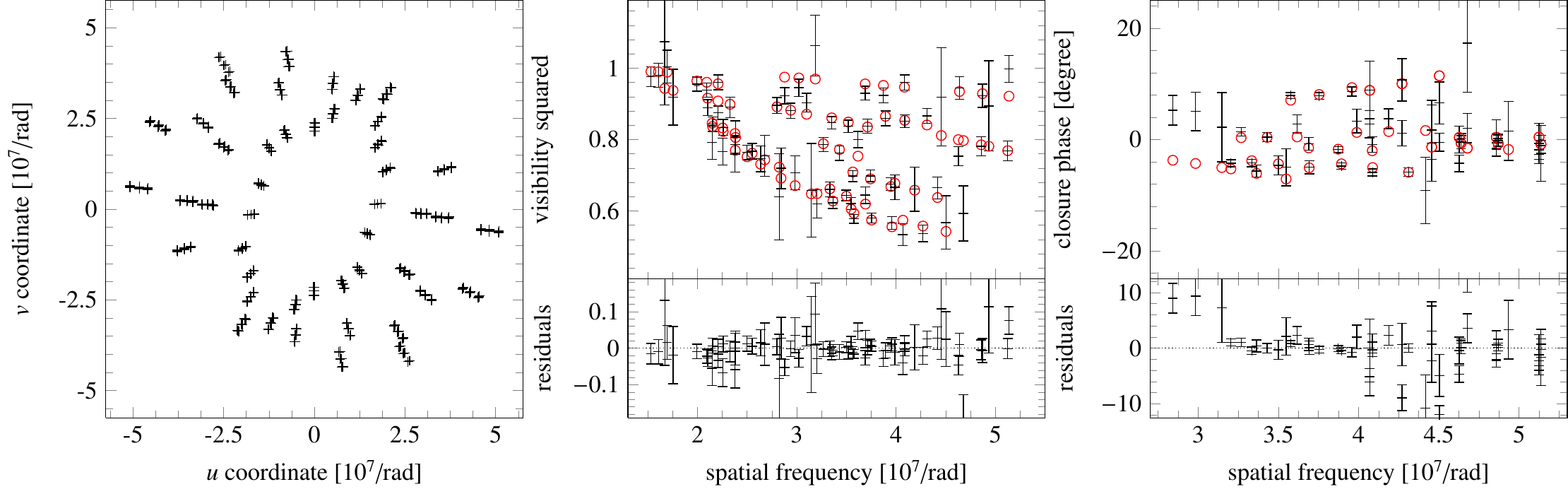}%

\caption{PIONIER observations of HR\,Car, showing $uv$-coverage, visibilities,
  and closure phases. Crosses mark the observed data, open circles the
  model, residuals are shown in the lower respective panels \citep[computed with LITpro, see][]{2008SPIE.7013E..1JT}.}
\label{Rivinius_HRCar_fig1}
\end{center}
\end{figure}

Analysis of the obtained PIONIER data firmly establishes the presence of two
different sources with a contrast ratio of about 1:5 and a separation of about
2mas (Rivinius et al., in prep.). Neither source is extended in itself, or
only marginally so, i.e., the data strongly support the binary hypothesis (see
Fig.~\ref{Rivinius_HRCar_fig1}).

Judging by the strength of the emission lines in various hydrogen lines and
the appearance of the visual spectrum in general, {a binary with a wind
  collision zone} seems to be the most attractive. While this interpretation
is work in progress, the pure {proof of binarity} was already delivered by the
PIONIER observations.

\subsection{Conclusions}

At the estimated distance of HR\,Car ($\sim$5000\,pc), observations indicate
that the two components have a projected separation of $\sim$10\,au. Assuming
a total mass of the system of $\sim 40$ M$_\odot$, this separation would
correspond roughly to an orbital period of about 5 years. A final value will
depend on the eccentricity and inclination of the system.  However, unless
HR\,Car is a wide system seen at a very unfavourable (and unlikely)
projection, {it would be the second known $\eta$\,Car-like LBV binary.} The
proposed scenario of a wind-wind effect being responsible for the AMBER
signature is similar to the model for the LBV candidate binary MWC\,314 by
\citet{2013A&A...559A..16L}.

\bibliographystyle{iau307}
\bibliography{HRCar}

\end{document}